\documentclass[aps,prb,floatfix,twocolumn,showpacs]{revtex4}
\usepackage{latexsym}
\usepackage{graphicx}
\newcommand{\figwidth}{3.275 in}
\usepackage{subfigure}    
\usepackage{epsfig}
\begin{document}
\title{Variational Monte-Carlo calculation of the nematic state of the 
two-dimensional electron gas in a magnetic field} 
\author{Quoc M. Doan$^{(1)}$ and Efstratios Manousakis$^{(1,2)}$ }
\affiliation{$^{1}$Department of Physics and MARTECH, 
Florida State University, 
Tallahassee, FL 32306-4350, USA and \\
$^{2}$Department of  Physics, University of Athens,
Penipistimiopolis, Zografos, 157 84 Athens, Greece.}  
\date{\today}
\begin{abstract}
We use a Jastrow-Slater wave function with an elliptical Fermi sea
to describe the nematic state of the two-dimensional electron gas
in a magnetic field and the Monte Carlo method to calculate a variational
energy upper bound. These energy upper bounds are compared with other
upper bounds describing stripe-ordered ground states which are
obtained from optimized Hartree-Fock calculations and with those which
correspond to an isotropic ground state. Our findings support
the conclusions drawn in our previous study where the Fermi-hypernetted
chain approximation was used instead of the Monte Carlo method.
Namely, the nematic state becomes energetically favorable relative
to the stripe-ordered Wigner crystal phase for the second excited
Landau level and below a critical value of the layer ``thickness'' 
parameter which is very close to its value in the actual materials. 
\end{abstract}
 \maketitle
\section{INTRODUCTION}
\label{chap:intro}
The  measurements  of  Lilly  et al. \cite{Lilly} and  Du et 
al.\cite{Du} reveal  strong anisotropic transport  properties  
of the two-dimensional electron gas (2DEG)  for the  half-filled 
Landau-level  system for high Landau-levels (LL) and at very low temperature.  
The  anisotropic behavior in the transport 
properties is consistent with stripe and bubble charge-density-wave 
phases which were predicted early on in 
Refs.~\onlinecite{Koulakov,Folger,Moessner} by means  of Hartree-Fock 
calculations  of the 2DEG and were confirmed more recently by 
numerical studies of systems with up to 12 electrons\cite{Haldane,Yang}.  
However, Fradkin et al.\cite{Kivelson} have 
challenged this interpretation and suggested that the anisotropic 
transport might be due to a possible nematic  
phase of the 2DEG in a magnetic field. This idea finds support in
the good comparison between the
results of the temperature dependence of the anisotropy of the
resistivity obtained by means of a Monte Carlo simulation 
of the nematic phase\cite{Fradkin} with that which has been
experimentally observed.
In addition, the idea is supported by the experiments of Cooper et 
al.\cite{Cooper} where an in-plane magnetic field  was applied  
in the 2DEG and the results of the experiment were interpreted 
on the basis of the presence of a nematic state; further support of
the idea is provided by the fact that
the theoretically estimated transition temperature from an isotropic
to nematic phase\cite{Wexler_Dorsey} is of similar magnitude as the 
experimentally determined temperature at which the on-set of the 
anisotropic transport occurs.

Rather recently we have presented\cite{FHNC_nematic} 
a variational calculation of the nematic state as ground state of the 
half-filled Landau-level system in a magnetic field based on an ansatz 
ground state wavefunction proposed by Oganesyan et al. \cite{Oganesyan} 
which is of the Jastrow-Slater form and is given by the following expression:
\begin{eqnarray}
\Psi \left( {\vec {r}_1  ,\vec {r}_2  ,...,\vec {r}_N } \right)&=&\hat
     {P}_0      \prod\limits_{j<k}^N     {(z_j     -}      z_k     )^2
     e^{-{\sum_{k=1}^N    |{z_k    } |^2}/4   } \nonumber \\
      &\times & \det   \left|   {\varphi   _{\vec   {k}}   (\vec   {r}_i   )}
     \right|\label{WF}
\end{eqnarray} 
where $\hat  P_0$  is the  projection operator   onto  the   lowest  LL,   
$\phi_{\vec  k}(\vec   r_i)$  are two-dimensional (2D) plane-wave states.  
Here, $z_j=x_j+iy_j$  is the complex 2D coordinate of the $j$ electron.  
This wavefunction  is a Jastrow correlated Slater determinant with Jastrow 
part similar to the Laughlin state \cite{Laughlin}.
This  ground-state wavefunction has the same  form as
the  form proposed by Rezayi and Read\cite{RR}, however, the
single-particle momenta form an elliptical Fermi sea as opposed to the
circular Fermi  sea. There is a  broken-symmetry  parameter which is
the  ratio $\alpha  =k_1 /k_2$  of the  semi-major $k_1$  and semi-minor
$k_2$ axes of the elliptic Fermi sea. Using this wave function to describe 
the nematic state we had carried out a  variational study of the half-filled 
system using the so-called  Fermi-hypernetted-chain (FHNC) 
approximation\cite{FHNC_nematic}.

The results of the above mentioned variational calculation indicate
that there is a certain value of the parameter $\lambda$ ($\lambda$ is
proportional to the 2DEG layer thickness\cite{ZDS}) below which the nematic
state is energetically favorable as compared to the isotropic and the
stripe-ordered ground states for the second excited LL. It is interesting
to note that this critical value of $\lambda$ is very close to the value 
of $\lambda$ which can be estimated based on the actual 
experimental conditions which are applicable for the case of the data by Lilly 
et al.\cite{Lilly} and by Du et al.\cite{Du}. However, one of the weak points
of the above described variational study is the fact that the FHNC
approximation is plagued by an unknown-size error and the results 
cannot be improved in a controlled manner\cite{Manousakis_Pandharipande}. 
Therefore, there is a need to check
the validity of these results and conclusions using the variational
Monte Carlo method and this task is undertaken in the present work.

There is a different variational approach to the problem of a broken
rotational state of the half-filled LL introduced by
Ciftja and Wexler\cite{Ciftja}. They have used the Fermi-hypernetted-chain
(FHNC) approximation to study a broken rotational state of the
half-filled LL where the symmetry-breaking parameter was introduced in
the correlation  part of the wavefunction as $(z_i-z_j)^2\rightarrow
(z_i-z_j-\alpha)(z_i-z_j+\alpha)$, and they used the standard
single-particle determinant with a circular  Fermi sea.

In this and in the  work of Ref.~\onlinecite{FHNC_nematic}, we 
considered the  unprojected wavefunction  of the
nematic state.  The  advantage of this simplified version  is that it
has a  Jastrow form  with a  Slater determinant so  it can  be applied
directly  with   FHNC and it allows  us  to  study  large-size 
 systems  using  the variational Monte Carlo method.
The paper is organized as follows: In the following Section
we discuss the formulation and the procedure; in Sec.~\ref{MC_Results}
we present the results and we compare them with those obtained
for the case of a stripe-ordered state and the isotropic state.
In Sec.~\ref{Conclusions} we summarize the conclusions of the
present calculation.

\section{Method}
\label{Procedure}

We  have adopted the toroidal geometry of a  square   with  periodic   boundary
conditions. This  geometry has the advantage of  naturally adapting to
the nematic and isotropic state wavefunction. 
There  are several steps  in applying
the MC  approach for  this problem. First,  as part of the  
wavefunction of nematic state we
construct a  Slater determinant of  plane waves  characterized by
momentum vectors which lie inside  
an  elliptical  Fermi  sea. Second,  since  the  pseudo-potential
is $ln(r)$,  which is a long-range  interaction, we need to
 to take  into
account all periodic image charge  interactions. One of the methods to
do this is the Ewald summation technique.
In   subsection \ref{Ewald} of the appendix  we  describe  the   
Ewald  summation technique for the case of toroidal boundary
conditions and the $ln(r)$ interaction. 
In the present section we will discuss our implementation of
the MC to study the nematic state. 

Given a value of $\alpha$ there are definite values of the number of
particles $N$ which  correspond to a closed shell.  These definite values
of $N$  are calculated as follows.  The  occupied states characterized
by $k_x,k_y$ must satisfy the following condition:
\begin{eqnarray}
\bigg(\frac{k_x}{k_1}\bigg)^2+\bigg(\frac{k_y}{k_2}\bigg)^2 \leq 1,
\end{eqnarray}
where $k_1$  and $k_2$ are the major  and minor axis of  the Fermi sea
and given by:
\begin{eqnarray}
k_1  =  \sqrt{\frac{4\pi   \rho}{\alpha}},\\  k_2  =  \sqrt{4\pi  \rho
  \alpha},
\end{eqnarray}
where $\rho$  is the  uniform particle density  of the system.   For a
finite system  of size  $L\times L$,  $k_x = n_x\Delta  k$ and  $k_y =
n_y\Delta k$ where  $\Delta k=2\pi/L$ and $n_x,n_y \in  Z$. So one can
deduce the conditions for $n_x,n_y$ such that:
\begin{eqnarray}
\frac{\pi}{N}\bigg(\alpha n_x^2+\frac{n_y^2}{\alpha}\bigg) \leq 1.
\label{n_x_n_y}
\end{eqnarray}
For  a value  of $N$  to be  acceptable the  number of states, i.e., the
number of pairs $(n_x,n_y)$ satisfying the above inequality should  
be equal to $N$.  For example,
for  $\alpha=1$,  $N$  can  be  $1,5,9,13,21,25,29,37,45,\ldots$;  for
$\alpha =2$, they can be $1,3,7,11,15,17,21,\ldots$
\begin{figure}[htp]
\vskip 0.3 in
\begin{tabular}{cc}
\epsfig{file=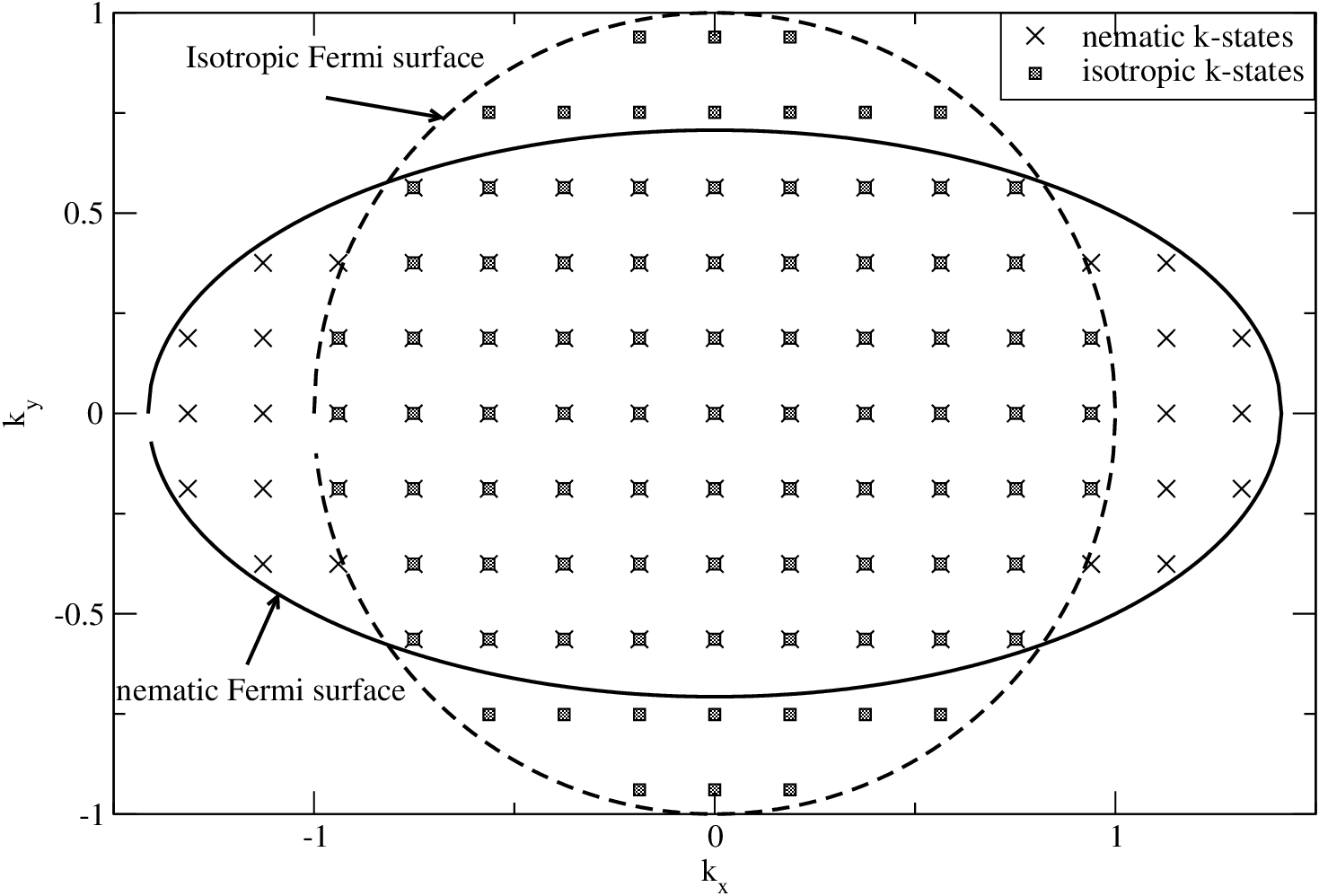,width=0.8\linewidth,clip=}\\
\epsfig{file=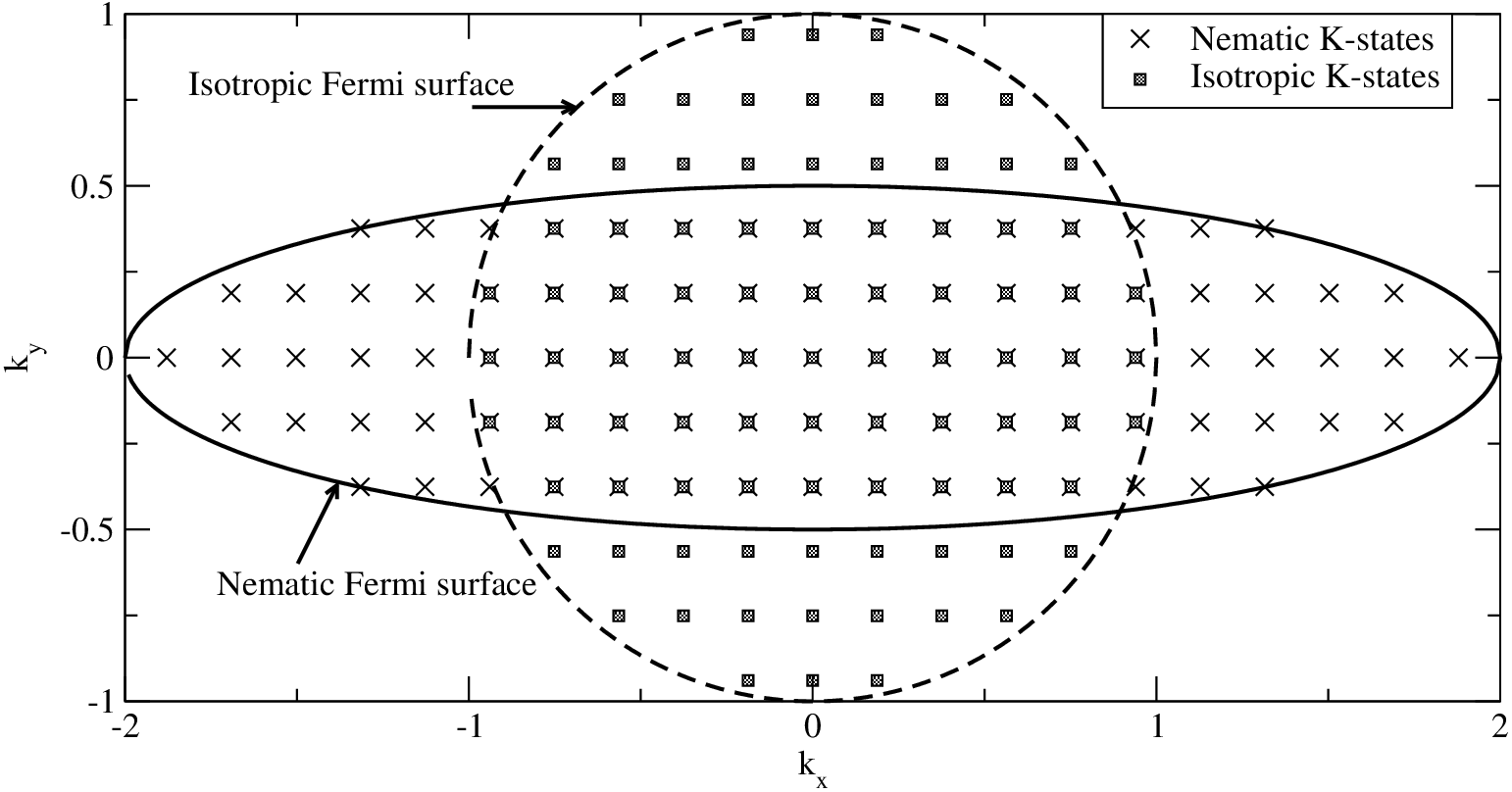,width=0.8\linewidth,clip=}
\end{tabular}
\caption{Occupied states for the nematic state with 
$\alpha=2$ (top) and $\alpha=4$ (bottom) for the case of 89 particles.}
\label{K89}
\vskip 0.3 in
\end{figure}

In Figs. \ref{K89}, we present two
examples  of  closed  shell  which  correspond to  $\alpha  =2$  and
$4$ for $N=89$. Notice that  with anisotropy parameter $\alpha=k_x/k_y  > 1$,
the occupied states (i.e  those satisfying equation \ref{n_x_n_y}) are
anisotropically distributed along the preferred $k_x$ axis.
In our  MC calculation we will use  these cases as well 
 as larger size systems up to  145 particles.  

We follow the  Metropolis MC scheme for
sampling the wavefunction where the ratio needed between the new and the 
old wavefunction is:
\begin{eqnarray}
\bigg \vert  \frac{\psi(\vec r_{new})}{\psi_(\vec r_{old})}\bigg \vert
^2= exp(u(\vec  r_{new})-u(\vec r_{old})) \bigg\vert\frac{Det(e^{i\vec
    k\cdot  \vec r_{new}})}{Det(e^{i\vec  k\cdot  \vec r_{old}})}\bigg
\vert.
\end{eqnarray}
where $u(\vec r)$ is the periodic pseudo-potential which is derived in
the subsection (\ref{Ewald}) of the appendix.
To  carry  out  the  calculation  of  the  ratio  between  the  Slater
determinant  of  the  new  configuration  and  the  old  configuration
efficiently,  we  use  the  inverse updating  technique  developed  by
Ceperley  et  al.\cite{Ceperley}.  We found that the number
of MC  steps needed for ``thermalization'' is  of the order of  $10^5$ and we
use of the  order of $2\times 10^6$ MC steps  to calculate averages of
the distribution function.

The potential energy of the high LL can be expressed \cite{Ciftja} via
the  pair distribution  function  of  the LLL  using  the single  mode
approximation discussed in Ref.~\onlinecite{MacDonald}, namely,
\begin{eqnarray}
V_L    =    \frac{\rho}{2}\int\bigg[g(\vec   r)-1\bigg]V_{eff}^L(r)d^2
r\label{V_L}
\end{eqnarray}
where  the effective potential  $V_{eff}^L(r)$ for Landau  level L  is the
convolution  of  the   effective  interaction\cite{ZDS}  
\begin{eqnarray}
V(r)  =  e^2
/\epsilon\sqrt{r^2+\lambda^2}
\label{lambda}
\end{eqnarray}
    with    the    $L$-order    Laguerre 
polynomial; namely, it is the Fourier transform of:
\begin{eqnarray}
\tilde{V}^L_{eff}(q)  =  \frac{2\pi  e^2}{\epsilon q}  e^{-\lambda  q}
\bigg[L_L(q^2/2)\bigg]^2\label{V_eff}
\end{eqnarray}
In the above formula, $\lambda$  is a length scale which characterizes
the  confinement  of  the  electron  wave function  in  the  direction
perpendicular to the heterojunction \cite{ZDS}.

We use the single mode approximation to calculate the interaction
energy at high LL (equation \ref{V_L}) and we are only interested in obtaining
the pair  distribution  function  $g(\vec r)$. The kinetic energy
advantage of the isotropic phase over the nematic phase is
calculated in the subsection (\ref{Kinetic}) pf the appendix.
The approach can be divided into the following steps:
\begin{itemize}
\item  The pair  distribution  function for  the LLL  for
  different anisotropic parameters $\alpha$ is calculated.
\item The single mode approximation\cite{MacDonald} is used to calculate
  the interaction energies at a high LL.
\item The kinetic  energy for different anisotropic parameters
  is evaluated (see appendix \ref{Kinetic}).
\item We compare total energies  of the isotropic and nematic state to
  determine at what LL the nematic becomes energetically favorable.
\item The optimum value of $\alpha$  is determined by minimizing the
total energy.
\item  The HF results which have been reported so 
far\cite{Stripe1,Stripe2,Stripe3} correspond to the case of 
$\lambda=0$.
Therefore, we needed to carry out Hartree-Fock calculations following 
Refs~\onlinecite{Stripe1,Stripe2,Stripe3} 
 for the case of the interaction given by Eq.~\ref{lambda} 
for $\lambda\ne 0$.
The  optimum total  energies  of  the  nematic states  will  be
  compared  with those  of the  stripe states  at different  values of
  $\lambda$ for the 2$^{nd}$ excited LL to determine  a critical value of
  $\lambda$ below which the nematic state maybe energetically favorable.
\item  The above mentioned critical  value of  $\lambda$  is  compared 
with  the value which corresponds to those samples used  in  the 
experiment\cite{Lilly}.
\item  A  comparison of  the  MC results  to the  ones
  obtained by FHNC will also be presented.
\end{itemize}

\section{Results}
\label{MC_Results}

\begin{figure}[htp]
\vskip 0.3 in
\includegraphics[width=\figwidth]{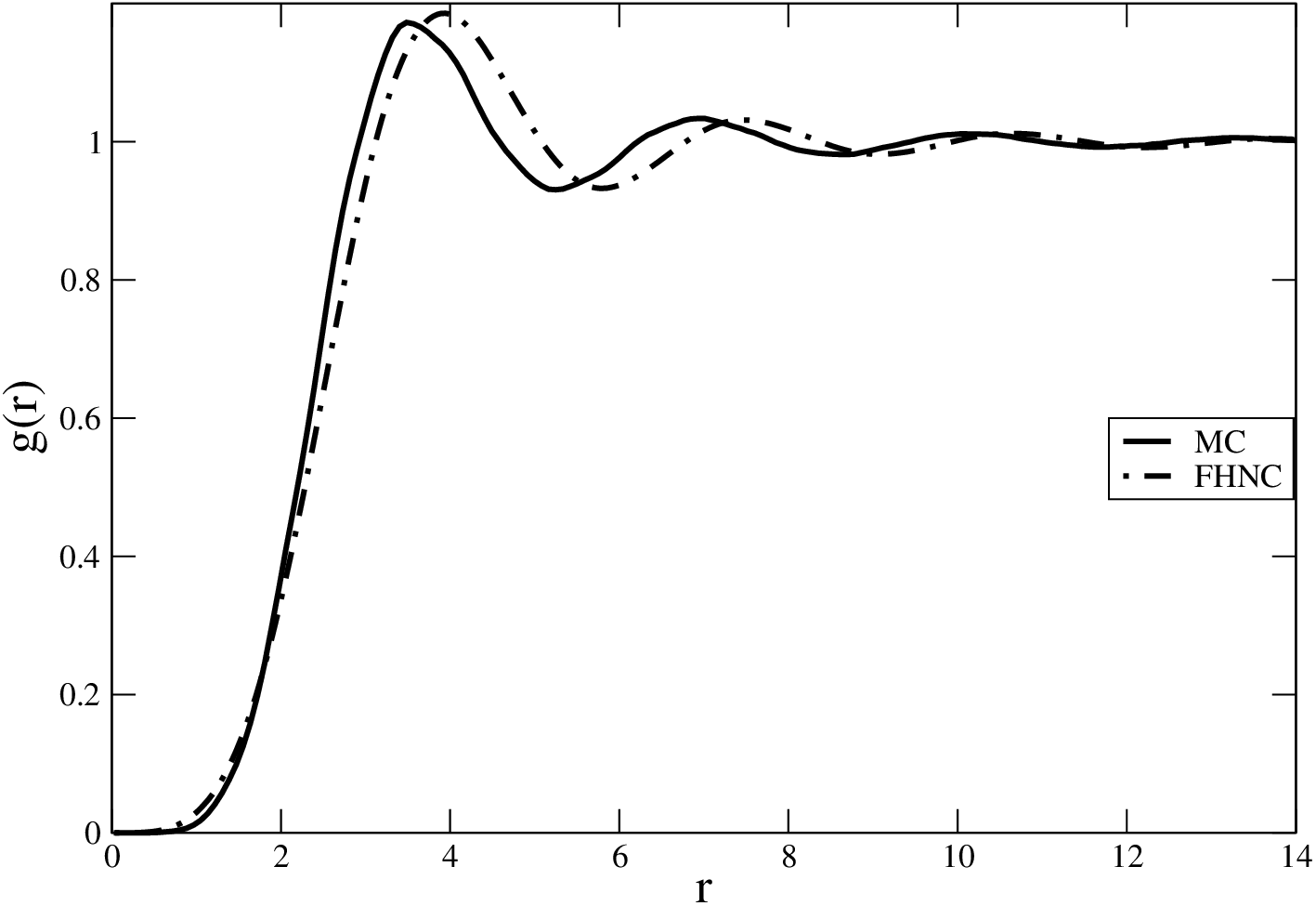}
\caption{Comparison of the pair distribution function obtained by FHNC
  and MC.}\label{g_FHNC_MC}
\end{figure}

The pair distribution  function $g(r)$ obtained
using MC integration has important differences  when compared to 
$g(r)$ obtained by
FHNC\cite{FHNC_nematic} as illustrated in Fig. \ref{g_FHNC_MC}. 
Thus, it is important to
obtain  the energies of  the nematic  state at  high LL  by MC  and to
compare them with those obtained by FHNC.

\begin{figure}[htp]
\vskip 0.3 in
\includegraphics[width=\figwidth]{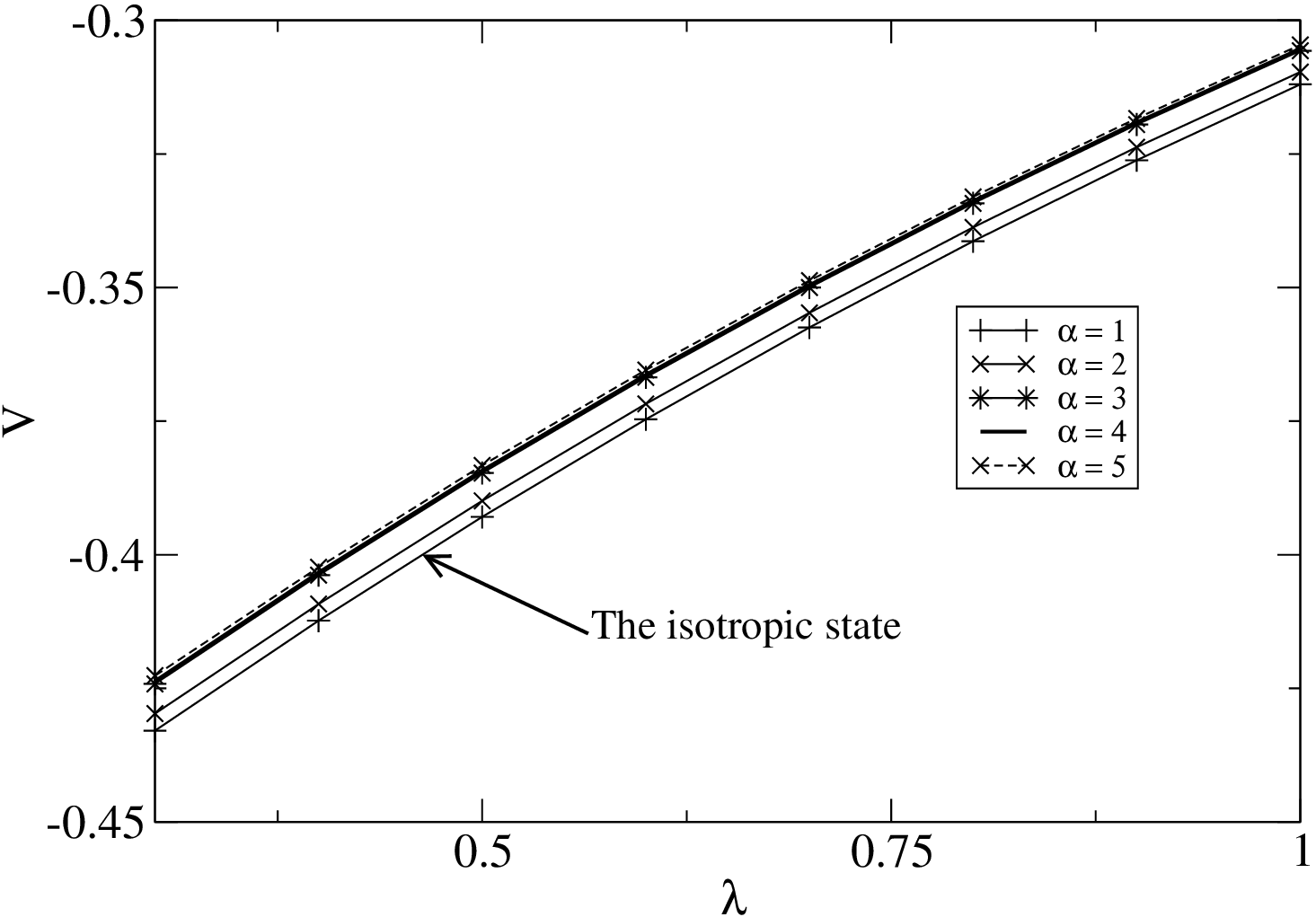}
\caption{Comparison  of  the potential  energy  of  the nematic  state
  calculated  for various  values of  $\alpha \neq  1$ as  function of
  $\lambda$    with    the    isotropic    state    ($\alpha=1$)  for
  LLL.}\label{V_0_MC}
\vskip 0.3 in
\end{figure}

\begin{figure}[htp]
\vskip 0.3 in
\includegraphics[width=\figwidth]{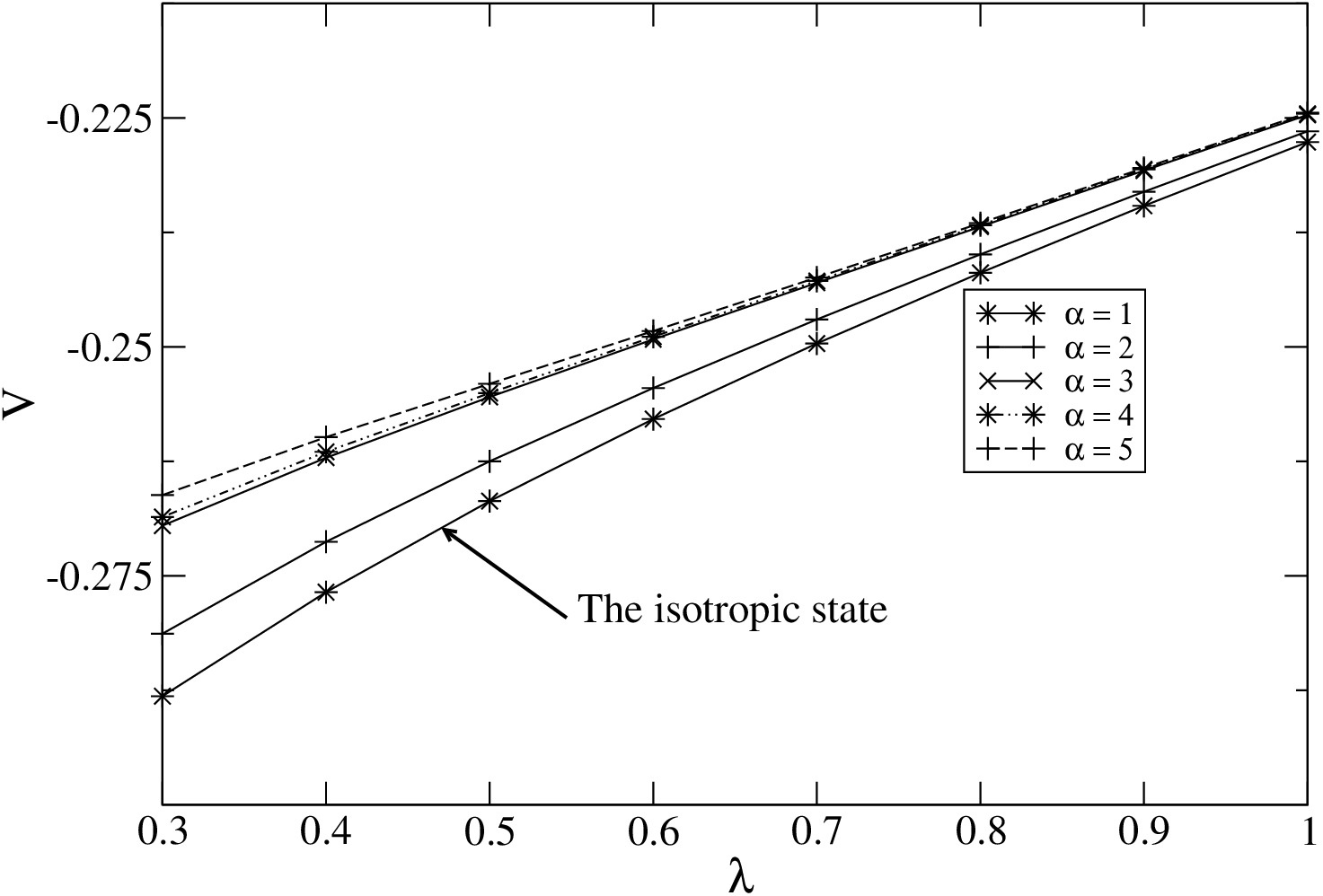}
\caption{Comparison  of  the potential  energy  of  the nematic  state
  calculated  for various  values of  $\alpha \neq  1$ as  function of
  $\lambda$  with the  isotropic  state ($\alpha=1$)  for the  1$^{st}$
  excited LL.}\label{V_1_MC}
\vskip 0.3 in
\end{figure}

We  first  compare  the interaction energies obtained for
different values  of $\alpha  > 1$ with  the potential energy 
of the  isotropic state
($\alpha=1$) (Figs. \ref{V_0_MC} and \ref{V_1_MC}).
Notice  from  Figs.  \ref{V_0_MC}  and  \ref{V_1_MC} that  the
potential energy of  the isotropic state is lower than the potential
energy of  the  nematic  state for  the  1$^{st}$  excited  LL and  LLL for
all values of the parameter $\lambda$.  The
potential  energy  is calculated  with  the pseudo-potential  obtained
using  the  Ewald  sum as discussed in subsection (\ref{Ewald}) of the
appendix.  Essentially the same  result is  also  found  with  a
pseudo-potential obtained using the Lekner summation 
technique\cite{Jensen}.  Furthermore, as shown in
appendix \ref{Kinetic},  the kinetic energy of the  isotropic state is
below that  of the nematic state,  and, thus, the total  energy of the
isotropic state is always lower than that of the nematic state. Hence,
our  MC   calculation  shows  that  the  isotropic   state  is 
energetically favorable  as compared to  the nematic state for the  LLL and
the  1$^{st}$ excited  LL for all values of the parameter $\lambda$. 
Also  note that the same conclusion was reached
using the FHNC technique\cite{FHNC_nematic} with the same wavefunction.
These  findings  solidify  the
conclusion that for the LLL  and the 1$^{st}$ excited LL, the isotropic
state  is  more  stable than  the  nematic  state,  which is  also  in
agreement  with the  experimental findings  of Refs.  \onlinecite{Lilly} and
\onlinecite{Du}.
\vskip 0.3 in
\begin{figure}[htp]
\vskip 0.3 in
\centerline{\includegraphics[width=\figwidth]{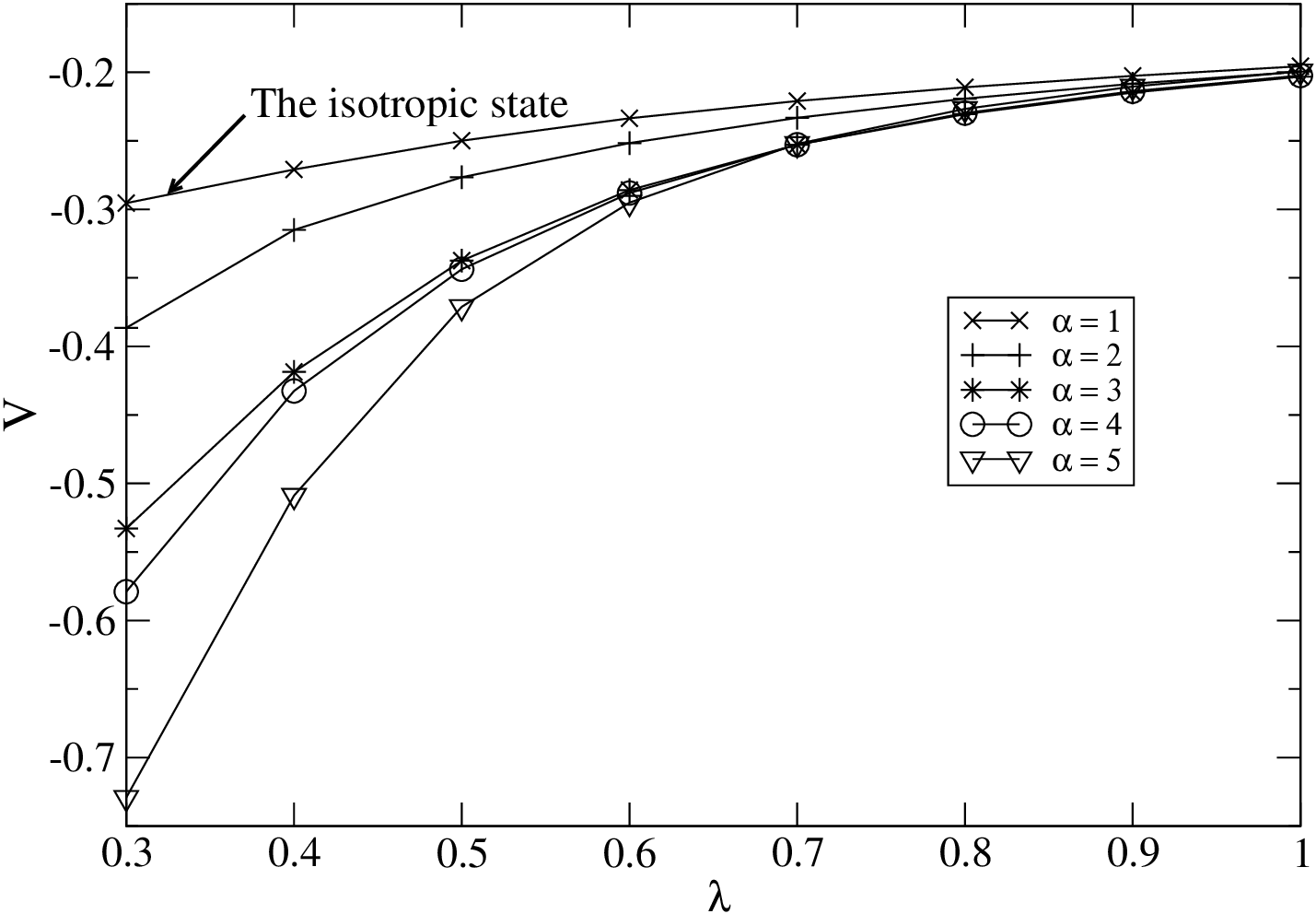}}
\caption{Comparison  between  the potential  energy  of  the nematic  state
  calculated for  various values of the  anisotropic parameter $\alpha
  \neq  1$   as a function  of  $\lambda$  and   the  potential 
energy of the isotropic  state
  ($\alpha=1$) at the $2^{nd}$ excited LL}\label{V_2_MC}
\vskip 0.3 in
\end{figure}
\begin{figure}[htp]
\vskip 0.3 in
\includegraphics[width=\figwidth]{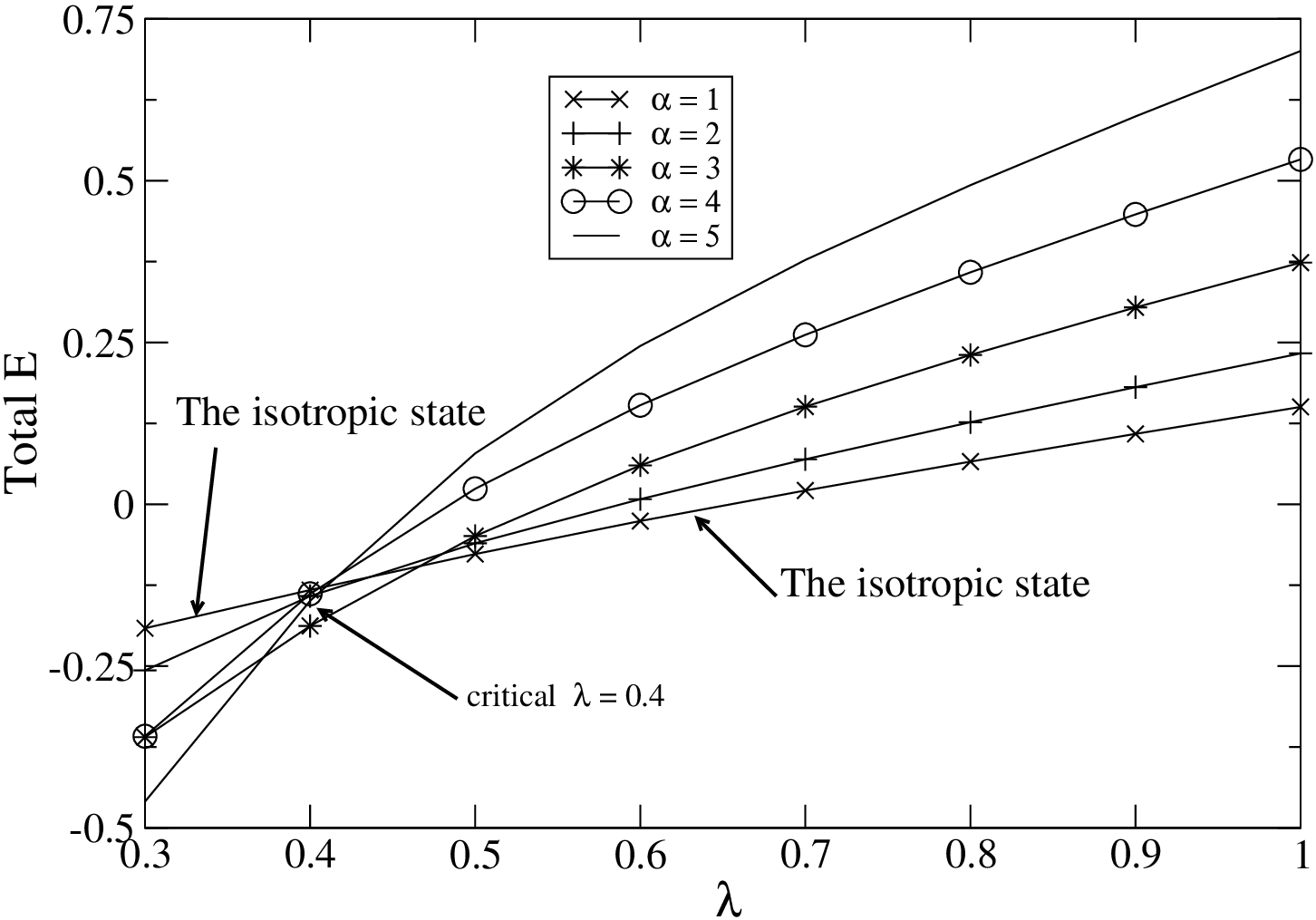}
\caption{Comparison of  total energy  of the nematic  state calculated
  for various values  of the anisotropic parameter $\alpha  \neq 1$ as
  functions of $\lambda$ with  the isotropic state ($\alpha=1$) at the
  $2^{nd}$ excited LL}\label{E_2_MC}
\vskip 0.3 in
\end{figure}
For  the  2$^{nd}$  excited  LL,  however, the  situation  changes  as
illustrated in Fig. \ref{V_2_MC}.
The conclusion which can be drawn from the comparison of 
Fig.~\ref{V_2_MC} is that the interaction energy of the nematic
state is lower than that of the  isotropic state for all values of 
$\lambda$. However, we need to
compare the total  energy of the nematic state with that of the 
isotropic state for the  2$^{nd}$ excited  LL (Fig. \ref{E_2_MC}).
From  Fig. \ref{E_2_MC}, we conclude that  the nematic
state is energetically favorable as compared to the isotropic state for
the  2$^{nd}$  excited LL  for  the range   of the parameter $\lambda  \leq
0.4$.  Note that using FHNC  we found\cite{FHNC_nematic}  that for 
$\lambda  \leq  0.6$ the total energy of the
nematic state is lower than the energy of the isotropic state. 
In summary, both
FHNC  and MC  yield similar  conclusions  about the  stability of  the
nematic state against the isotropic state for the 2$^{nd}$ excited LL.

\begin{figure}[htp]
\vskip 0.3 in
\includegraphics[width=\figwidth]{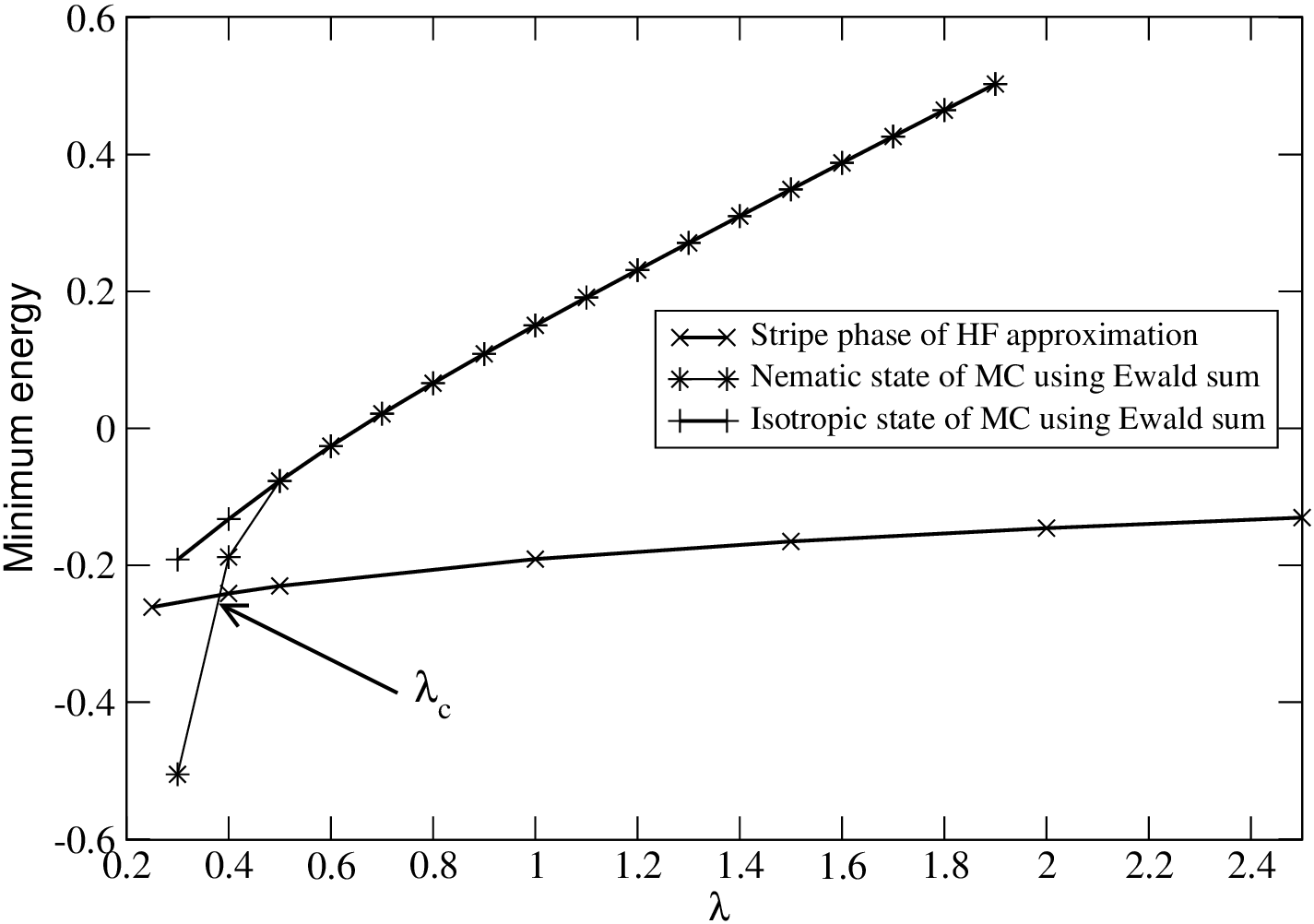}
\caption{Comparison  of the  optimal  nematic state  calculated by  MC
  using the pseudo-potential using the Ewald sum with the stripe state
  calculated by HF as function of $\lambda$.}\label{Ewald_MC_HF}
\vskip 0.3 in
\end{figure}

In Refs.~\onlinecite{Koulakov,Folger,Moessner} 
the stripe-ordered phase was predicted based on HF calculations
and this ordering can also explain the anisotropy
observed   in   the   transport    properties   of   the   2DEG   at   low
temperature. Therefore,  we need to investigate  the stability of
the nematic  state against the stripe-ordered state  as follows. First,
we find the optimum energies of  the nematic state with respect to the
anisotropic parameter $\alpha$ for various values of $\lambda$. Next,
we  compare  these with  the  optimum  energies  of the  stripe  state
obtained by the HF approximation\cite{Stripe1,Stripe2,Stripe3}.
Calculations for  the case  where
$\lambda=0$  have  been  carried   out  in  Refs.  \onlinecite{Stripe1,Stripe2}
and \onlinecite{Stripe3}. For making a comparison with the optimum
energy of the nematic state at various values of $\lambda$, 
we carried out detailed HF  calculations for the  case where 
$\lambda \neq  0$.   For the  stripe-ordered state, the
optimum energy  is obtained by  minimizing the energy with  respect to
the uniaxial anisotropy parameter  $\varepsilon$ defined in 
Ref.~\onlinecite{Stripe3}.  
Fig. \ref{Ewald_MC_HF} shows the comparison of the optimal
energies obtained  by MC  for the nematic  state with the optimum (with 
respect to $\varepsilon$) energy for the  stripe state
obtained by HF. Note that, for $\lambda \geq 0.5$, the optimum nematic
state is obtained for $\alpha=1$, i.e., it is the isotropic state.
Furthermore, Fig. \ref{Ewald_MC_HF} demonstrates that  the nematic  state is
energetically lower than  the stripe state for the  values of $\lambda
\leq \lambda_c  = 0.37$.     
As discussed earlier the
pseudo-potential  can be  obtained by  using either  the Ewald  or the
Lekner    summation   technique\cite{Jensen}.
We have also carried out the same calculation using 
the  Lekner summation technique and the results obtained
are in good agreement with those obtained using the Ewald  summation
method.  Thus, we can conclude
that  with  MC calculation,  for  $\lambda  \leq \lambda_c=0.37$,  the
energy of  the nematic state  is lower than the  stripe-ordered state.
\section{Conclusions}
\label{Conclusions}
\begin{figure}[htp]
\vskip 0.3 in
\includegraphics[width=\figwidth]{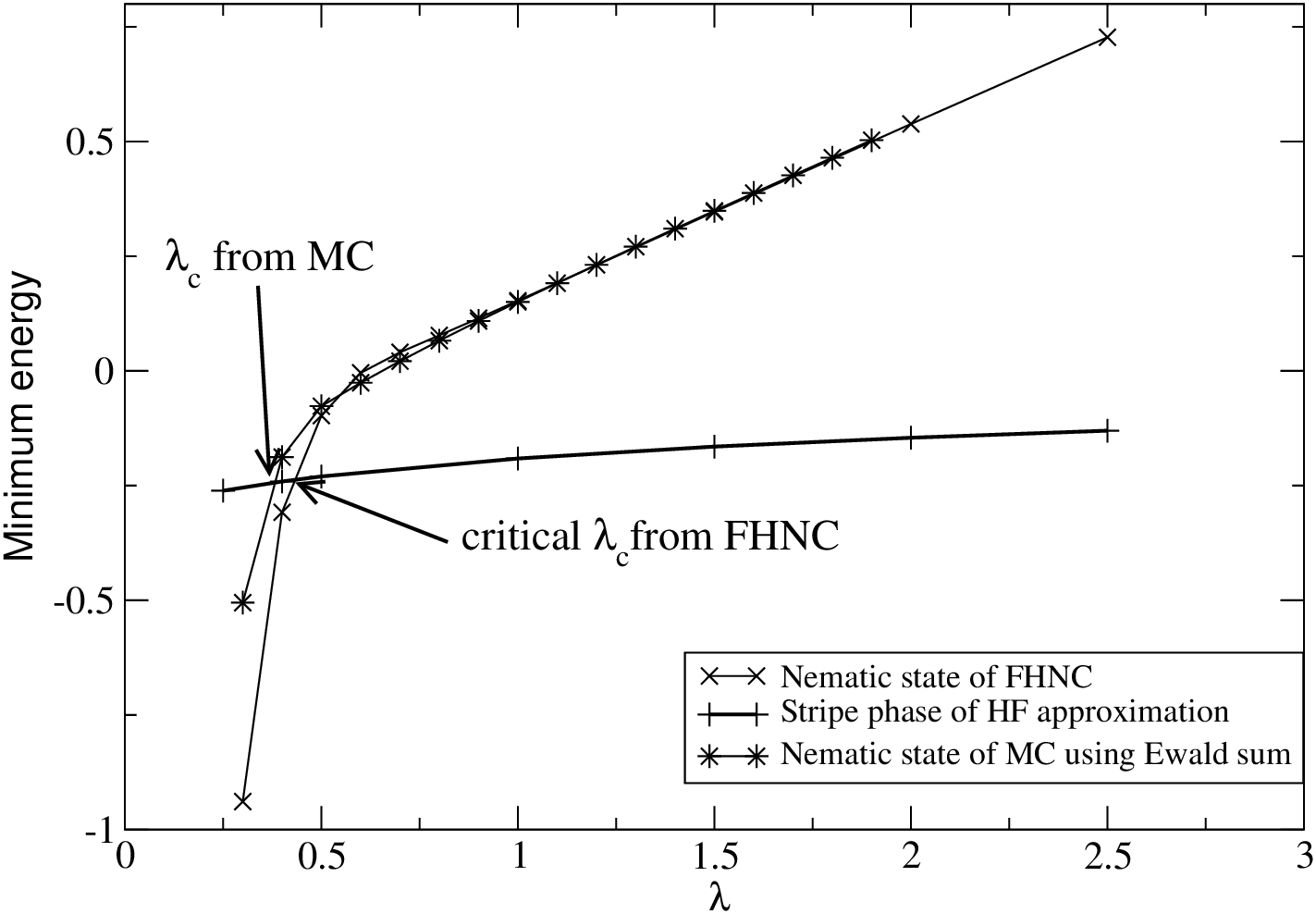}
\caption{Comparison of  the optimum  nematic state obtained  from FHNC
  and MC with the stripe state obtained from HF}\label{Ewald_MC_FHNC}
\vskip 0.3 in
\end{figure}

In Fig.~\ref{Ewald_MC_FHNC} the results for the optimum 
total energy of the nematic state obtained with the variational MC
method is compared with that obtained by FHNC in Ref.~\onlinecite{FHNC_nematic}
and with the optimum energy of the stripe-ordered state. 
The critical  value of $\lambda_c$ we found  from FHNC\cite{FHNC_nematic} 
is $0.4$ which    is    close to the value of 0.37 obtained above by MC.  
The critical  value of $\lambda$  corresponding to the sample  used in
experiment \cite{Lilly} which was calculated in 
Ref.~\onlinecite{FHNC_nematic}, using the conditions of the experiment
and sample characteristics,
is approximately  $0.34$, which  can be  below the  critical  value found
above. Thus,   both  MC  and  FHNC  calculations
indicate  that   the  nematic  state  might  be   the  state  observed
experimentally for the 2DEG at  the heterojunction in the samples used
in experiment described in Ref.\onlinecite{Lilly}.
There is still a remaining question about the validity of our approximation
to neglect the projection operator in the wavefunction (\ref{WF}).
However, in both FHNC treatments of the problem\cite{FHNC_nematic,Ciftja}, 
where, in addition to neglecting the projection operator for arguments
presented there, there was a second rather annoying question (and rather
straightforward to answer) of 
the validity of the FHNC approximation in evaluating the energy
expectation value. In the present paper the latter question is answered
by employing the Monte Carlo method.
Therefore, we conclude that the present calculations eliminates 
the suspicion that the conclusions drawn in Ref.~\onlinecite{FHNC_nematic}
might be due to an artifact of the FHNC approximation employed in 
Ref.~\onlinecite{FHNC_nematic}.

\section{Appendix}

\subsection{Ewald summation technique for the logarithmic potential}
\label{Ewald}

The long-range nature of the pseudo-potential $ln(r)$ which appears in
the exponent of the wavefunction of the Jastrow factor in the case of
periodic boundary conditions requires a summation over all periodic 
image charges. Specifically, the ``charge'' distribution required to
give rise to a logarithmic interaction is given as:
\begin{eqnarray}
\rho(\vec r) = \sum_{\vec R}\delta(\vec r-\vec R)+\rho_{background}
\end{eqnarray}
The two-dimensional (2D) Poisson equation is given by:
\begin{eqnarray}
\nabla^2\Phi(\vec r) = -2\pi \rho(\vec r)
\end{eqnarray}
and its solution in 2D  is the logarithmic interaction.
We need to solve the above equation for a periodic square $L \times L$.
The idea of  Ewald summation is to add around  each charge an opposite
Gaussian charge  distribution of an appropriately  chosen width $\mu$,
and,   in   addition,   to   subtract   the   same   Gaussian   charge
distribution.    
Let us  split $\rho$ into  long-range and short-range portions  in the
following manner:
\begin{eqnarray}
\rho(\vec   r)  =   \rho_1(\vec  r)   +\rho_2(\vec   r)\\  \rho_1(\vec
r)=\sum_{\vec              R}\frac{e^{-\frac{(\vec              r-\vec
      R)^2}{\mu^2}}}{\pi\mu^2}+\rho_{background}\\          \rho_2(\vec
r)=\sum_{\vec   R}\bigg[\delta(\vec   r-\vec  R)-\frac{e^{-\frac{(\vec
        r-\vec R)^2}{\mu^2}}}{\pi\mu^2}\bigg]
\end{eqnarray}
$\phi_1$, which corresponds to $\rho_1$ is a  short-range potential, 
and,  thus, we can calculate  $\phi_1$ in
real space since it converges very quickly.  The other combined charge
configuration, i.e., $\rho_2$, consisting  of the Gaussian and  the 
background charge and the corresponding potential  is denoted by
$\phi_2$. Since $\phi_2$  is a long-range potential it will be calculated
in Fourier  space. The solution to each of the Poisson's equations
for the two charge distributions and the corresponding potential is
straightforward. We note that for our  case the ``charge'' of  the particle 
is $e^2=2m$. We find 
\begin{eqnarray}
\phi_1(\vec             r)=\frac{4m\pi}{A}\sum_{\vec             k\neq
  0}\frac{e^{-\mu^2k^2/4}}{k^2}e^{i\vec                      k\cdot\vec
  r}\label{Ewald_1},\\           \phi_2(\vec           r)=-m\sum_{\vec
  R}Ei\bigg[-\frac{(\vec r-\vec R)^2}{\mu^2}\bigg]\label{Ewald_2}.
\end{eqnarray}
where $\vec k = 2\pi/L \vec n$ with $\vec n\in Z^2$ and 
$Ei(t)$ is the Exponential  integral function and is defined by:
$Ei(t)  = -\int_{-t}^{\infty}\frac{e^{-x}}{x}dx$.

For  the  Ewald  summation,  the convergence  of  (\ref{Ewald_1})  and
(\ref{Ewald_2}) is achieved choosing  the width of the Gaussian charge
distribution   $\mu=1$,  the   number  of   cells  for   the   sum  in
(\ref{Ewald_2}) to be 10 and by carrying out the sum in momentum space
in (\ref{Ewald_1})  over 200 k-states.

In order to check the validity of this approach for the case
of our use of toroidal boundary conditions we calculated the
distribution function and the energy for the 1/3 ($m=3$) case
using the expressions (\ref{Ewald_1},\ref{Ewald_2}) and our
results for the energy and distribution function are identical
to the results of Morf and Halperin\cite{Morf} who
used the disk geometry.

\subsection{Evaluation    of    kinetic    energy   of    the    nematic
  state}\label{Kinetic}  
In  this subsection of the appendix,   we compute the
kinetic energy difference between the nematic and the
isotropic state. In the single-LL approximation, the kinetic energy is
quenched. In  addition, the same  is true in  the HF treatment  of the
stripe,  namely, there  is no  kinetic energy  due to  any correlation
factors  or operators.  While this  approximation gives  a significant
difference between the potential energy of the isotropic state and the
nematic  state, it gives  no difference  between their  kinetic energy
which is unacceptable because of the difference in the geometry of the
Fermi sea. We want to estimate this difference. We can start with:
\begin{eqnarray}
  (\vec \nabla-\vec A)^2F\Phi\\
=     (\vec     \nabla-\vec     A)^2F\Phi+2\bigg[(\vec     \nabla-\vec
  A)F\bigg]\nabla\Phi+F\nabla^2\Phi
\end{eqnarray}
The first term in the above equation yields:
\begin{eqnarray}
(\vec \nabla-\vec A)^2F\Phi = \frac{\hbar\omega_c}{2}F\Phi,
\end{eqnarray}
which  is  common  for  all  states under  our  consideration  so  for
simplicity we can drop it.  The last term is:
\begin{eqnarray}
F\nabla^2\Phi=F\sum_{k}\frac{\hbar^2k^2}{2m^{\star}}\Phi
\end{eqnarray}
So the contribution of the last term is:
\begin{equation}
\sum_{\vec k \in FFS}\frac{\hbar^2k^2}{2m^{\star}}
\end{equation}
where $\vec k \in FFS$ stands for a summation over all vectors
$\vec k$ in the corresponding filled Fermi sea.
The summation over the circular Fermi sea in the isotropic case
 is given by:
\begin{eqnarray}
\frac{1}{N}\sum_{\vec k}\frac{\hbar^2
  k^2}{2m^\star}=\frac{\hbar^2k_F^2}{4 m^\star},
\end{eqnarray}
and the in the case of the elliptic Fermi sea in the anisotropic 
case the summation is given by
\begin{eqnarray}
\frac{1}{N}\sum_{\vec k}
\frac{\hbar^2k^2}{2m^\star}=\frac{\hbar^2}{4m^{\star}}\frac{k_1^2+k_2^2}{2}
\end{eqnarray}
Using the  facts that  $k_F^2=k_1\cdot k_2$ and  $k_1/k_2=\alpha$, the
kinetic energy difference between  the isotropic state and the nematic
state is given as follows:
\begin{eqnarray}
\Delta (KE) = -\frac{\hbar^2k_F^2}{4m^{\star}}\frac{(1-\alpha)^2}{2\alpha}.
\end{eqnarray}

\section{Acknowledgements}
We would like to thank Eduardo Fradkin, Steve Kivelson and Kun Yang
and Lloyd Engel for useful discussions.

\end{document}